\documentclass{ws-procs961x669}            

\newcommand{\Dk}[1]{\frac{d^3#1}{(2\pi)^3}}

\newcommand{\ve}[1]{{\text{\bf #1}}} 
\newcommand{\vk}{\ve k}
\newcommand{\vp}{\ve p}
\newcommand{\vq}{\ve q}
\newcommand{\vx}{\ve x}

\newcommand{\vr}{\ve r}

\newcommand{\dD}{\delta_\text{D}}

\newcommand{\citar}[1]{\textcolor{red}{CITAR}}

\begin{document}
\title{Testing modified gravity theories with marked statistics}

\author{Alejandro Aviles}

\address{Consejo Nacional de Ciencia y Tecnolog\'ia,\\
Av. Insurgentes Sur 1582,
Colonia Cr\'edito Constructor, Del. Benito Ju\'arez, 03940, Ciudad de M\'exico, M\'exico, \\
and \\
Departamento de F\'isica, Instituto Nacional de Investigaciones Nucleares \\
Apartado Postal 18-1027, Col. Escand\'on, Ciudad de M\'exico,11801, M\'exico.\\
E-mail: avilescervantes@gmail.com}

\begin{abstract}
In the last two decades, Modified Gravity (MG) models have been proposed to explain the accelerated expansion of the Universe. However, one of the main difficulties these theories face is that they must reduce to General Relativity (GR) at sufficiently high energy densities, such as those found in the solar system. To achieve this, MG theories typically employ so-called screening mechanisms: nonlinear effects that bring them to GR at the appropriate limits. For this reason, low-energy regions where the screenings do not operate efficiently, such as cosmic voids, are identified as ideal laboratories for testing GR. Hence, the use of {\it marked} statistics that up-weight low energy densities have been proposed for being implemented with data from future galaxy surveys. In this proceeding note, we show how to construct theoretical templates for such statistics and test their accuracy with the use of N-body simulations.
\end{abstract}


\bodymatter

\section{Modified gravity in Cosmology: a very brief introduction}

The standard model of Cosmology, the so-called $\Lambda$CDM, is the most successful theory to date for describing the large-scale behaviour of our Universe. The outstanding way this model fits the cosmic background radiation (CMB) anisotropies measurements from the Planck satellite\cite{Aghanim:2018eyx}---and prior to that from the WMAP and COBE experiments---leaves small room for modifications of it. However, among other drawbacks in Cosmology, nowadays cosmic acceleration is still not well understood, and in models that modify the background history of the Universe at late times, the CMB differs mainly because the journey of photons from the last-scattering surface to us is affected. This is a secondary effect called the integrated Sachs-Wolfe effect and its influence over the CMB is mainly localized at small multipoles, where the error on measurements is nonetheless large because of the cosmic variance---the fact that we have only one universe and hence a small number of long wave-length modes to average over. Such a liberty has led to a lot of proposals (literally hundreds of them) as alternatives to the cosmological constant to explain the speeding up of the expansion of the Universe. Unfortunately, there is much more literature proposing new gravity models or that makes simple tests to them (the vast majority focusing only on the homogeneous and isotropic cosmology; I will come back to this in a moment), than literature that explore methods to probe the clustering due to gravity at cosmological scales. Of course, this is a highly unbalanced situation and observational cosmologists cannot test each of the proposed models. These alternatives can be categorized into two big branches: dark energy (DE) and Modified Gravity (MG); though, one should be careful since there is not a well defined boundary between them. In this proceedings note, we will focus on the latter assuming well defined, and to the particular case of chameleon fields\cite{Khoury:2003aq,Khoury:2003rn} although the techniques described here have been applied to other kinds of theories.

\begin{figure}
	\begin{center}
\includegraphics[width=3.2in]{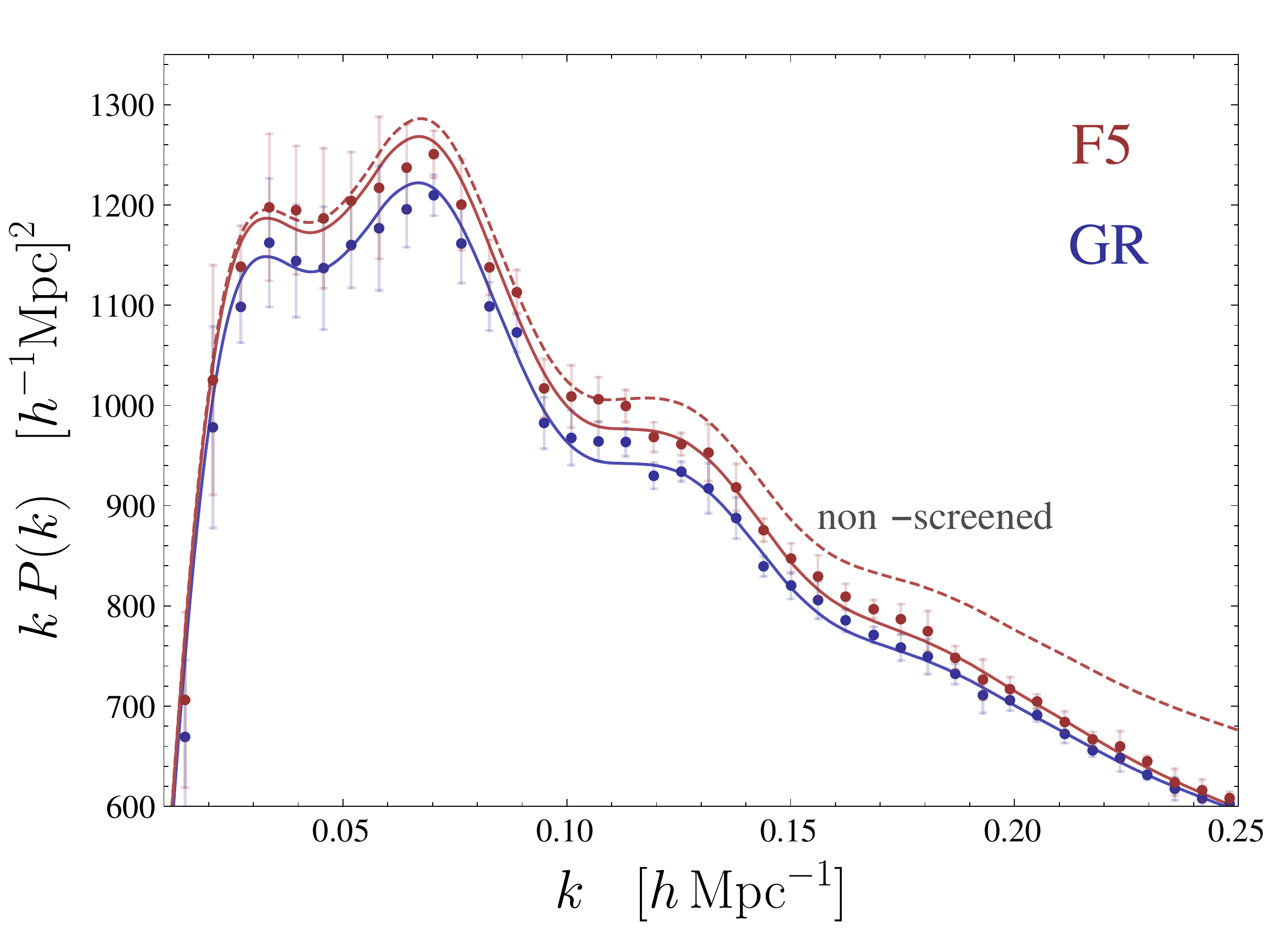}
\caption{Matter power spectrum in the models GR (blue line) and Hu-Sawicki F5 (red lines). The latter is further shown with and without screenings with solid and dashed lines, respectively. To generate this analytical results we use the Standard Perturbation Theory with effective field theory counterterms and infrared-resummations of Refs.~\citenum{Aviles:2017aor,Aviles:2020wme}. The mock data is extracted from the \texttt{Elephant} simulations.}
\label{fig:chameleon}
    \end{center}
\end{figure}

Perhaps the most difficult aspect of modifying gravity is the following. Observations indicate that the {\it true} model of the universe should be close to $\Lambda$CDM, at least from the epoch of the primordial (``Big-Bang'') Nucleosynthesis 
until now. In particular the expansion history should be close to that of $\Lambda$CDM, perhaps with small differences with the aim of, for example, resolving the Hubble tension.\cite{Verde:2019ivm}
However, as a rule of thumb, MG models that slightly depart from $\Lambda$CDM at the background level, become quite different at the perturbative level.\cite{Bertschinger:2006aw,Bertschinger:2008zb} That is, two models that have almost the same expansion history, typically predict a completely different clustering. This is because in MG the two scalar gravitational potentials of the metric, thinking of in Newtonian gauge, are different even in the absence of anisotropic stresses, which makes GR a very special theory and not easy to mimic. This would mean that the gravitational potential appearing in the geodesic potential is not the same to the gravitational potential sourced by the energy density fields in the Poisson equation; and what is particularly important to this discussion is that the homogeneous and isotropic expansion is completely insensitive to this. For example, take a look at figure \ref{fig:chameleon}, where we show the power spectrum of matter density fluctuations in GR and in the F5 gravitational model (we will define what is F5 in appendix \ref{app:HS}). The F5 gravitational model is indistinguishable for any practical purpose from GR at the background level, but their predicted matter power spectra differ by a lot. This is the reason why it does not make too much sense to test MG models at the background level only, for example using supernovae data only. However, for the sake of fairness we must say there are exceptions to this rule, for example the recently proposed No-Slip gravity,\cite{Linder:2018jil} tailored with the specific purpose that the two scalar gravitational potentials become equal.

When gravity is modified, in particular in the infrared, one should be careful to not spoil the well-tested regimes of validity of GR, as for example in the solar system or regions with much larger densities,\footnote{It is perhaps surprising to consider the solar system a high energy density region, but it is very large compared to the mean background energy density $3 H_0^2 / 8 \pi G \sim 10^{-29}\,\text{g}/\text{cm}^3$. The latter is the one that is important at the largest cosmological scales, and hence drives the background history of the Universe. }  but at the same time provide the accelerated expansion of the Universe. In order to deal with this unquestionable observational constriction, MG models often invoke mechanisms that yield the theory to GR in the appropriate limits. These are generically called {\it screening} mechanisms: non-linear effects that drive MG theories to GR when either the environmental energy density is large, or there are large gravitational tidal fields, among other scenarios. The most popular is the {\it chameleon} mechanism,\cite{Khoury:2003aq,Khoury:2003rn} in which a scalar field mediates a universal fifth force, but its mass depends on the environmental density where it resides: as larger is the ambient density, the lighter is the mass of the scalar field and smaller the range of the force. For example, in successful theories, and for human-like densities ($\sim$ 1 gram per cubic centimeter) the range of the fifth-force becomes so ridiculously small that it becomes effectively invisible to any experiment. On the other hand, at extremely low ambient densities the range of the fifth-force becomes quite large, with cosmological scope.

The most studied MG theory in Cosmology, one can affirm with high confidence, is the $f(R)$ gravity,\cite{Starobinsky:1980te,Carroll:2003wy,Capozziello:2002rd} and in particular the Hu-Sawicki model.\cite{Hu:2007nk} Although, these theories were first understood as higher-order gravity theories, it was soon recognized that under a few assumptions
they can be described as second order differential equations for the metric plus a scalar field\cite{Magnano:1993bd,Jaime:2010kn} which propagates only an additional massive mode. In this proceeding note we will lead with the Hu-Sawicki model to exemplify our analytical results and to take advantage of the fact that we have good, {\it state-of-the-part} N-body simulations for them---the Extended LEnsing PHysics using ANalaytic ray Tracing \texttt{Elephant} cosmological suite of simulations.\cite{2012JCAP...01..051L,Cautun:2017tkc}  Even though $f(R)$ is not a representative model of the vast universe of MG models (or more narrowly of Hordenski theories), the techniques reviewed here can be applied to different models and can be beneficial to study MG that rely on screening mechanisms. The material presented in this note is extracted mainly from Refs.~\citenum{White:2016yhs,Aviles:2019fli,Alam:2020jdv,Philcox:2020srd}; we refer the reader to these works for further details.

\section{The importance of up-weight low energy density regions}

In figure \ref{fig:chameleon} we show the chameleon screening in action. The black line is the analytical power spectrum of the $\Lambda$CDM. The red coloured lines show the matter power spectrum in the Hu-Sawicki F5 model with screenings (solid line) and without screenings (dashed line), together with data from the \texttt{Elephant} simulations. We note the screening mechanism drives the theory to GR, more evident at small scales (high wave-numbers $k$), while without the screenings the differences between the two models are larger. This behaviour raises the question if the power spectrum, and its counterpart in configuration space, the correlation function, are the most clever way to test the gravitational theories from its cosmological clustering. After all, by construction these statistics give more weight to high density regions, since they sum over all tracers evenly (either particles, halos, galaxies...), and tracers tend to clump together by gravitational attraction. Meanwhile, cosmological voids are underrepresented by the power spectrum---as well as by the correlation function and their N-points generalizations---even though they are the largest regions in our universe, and as such, they contain a significant share of the total energy budget.  But clumpy regions of space, being denser, behave very close to GR due to the screenings mechanism. Hence, the standard statistics used in Cosmology may not be the best option to test infrared modifications of GR. 

With the above ideas in mind, M.~White proposed in Ref.~\citenum{White:2016yhs} to use {\it marked correlation functions}\footnote{Marked statistics have a long history;\cite{Beisbart:2000ja} they have been used to assign properties to objects, such as the luminosity, color, and morphology of galaxies,\cite{Sheth:2005ai,Sheth:2005aj} and to break degeneracies between Halo Occupation Distribution (HOD; a semi-analytical method to populate halos from N-body simulations with galaxies) and cosmologies.\cite{White:2008ii} 
} to test gravity at cosmological scales. The proposal consists in weighting the tracer field density fluctuations with a function (the mark) whose value depends on the environmental density, such that objects residing in low density regions become up-weighted in 2-point statistics, while objects residing in high density become down-weighted. In this way, statistics of such marked density fields will probe low density regions where gravitational fifth-force screenings are inefficient and the effects of hypothetical MG theories would be more pronounced.

\begin{figure}
	\begin{center}
\includegraphics[width=5.5in]{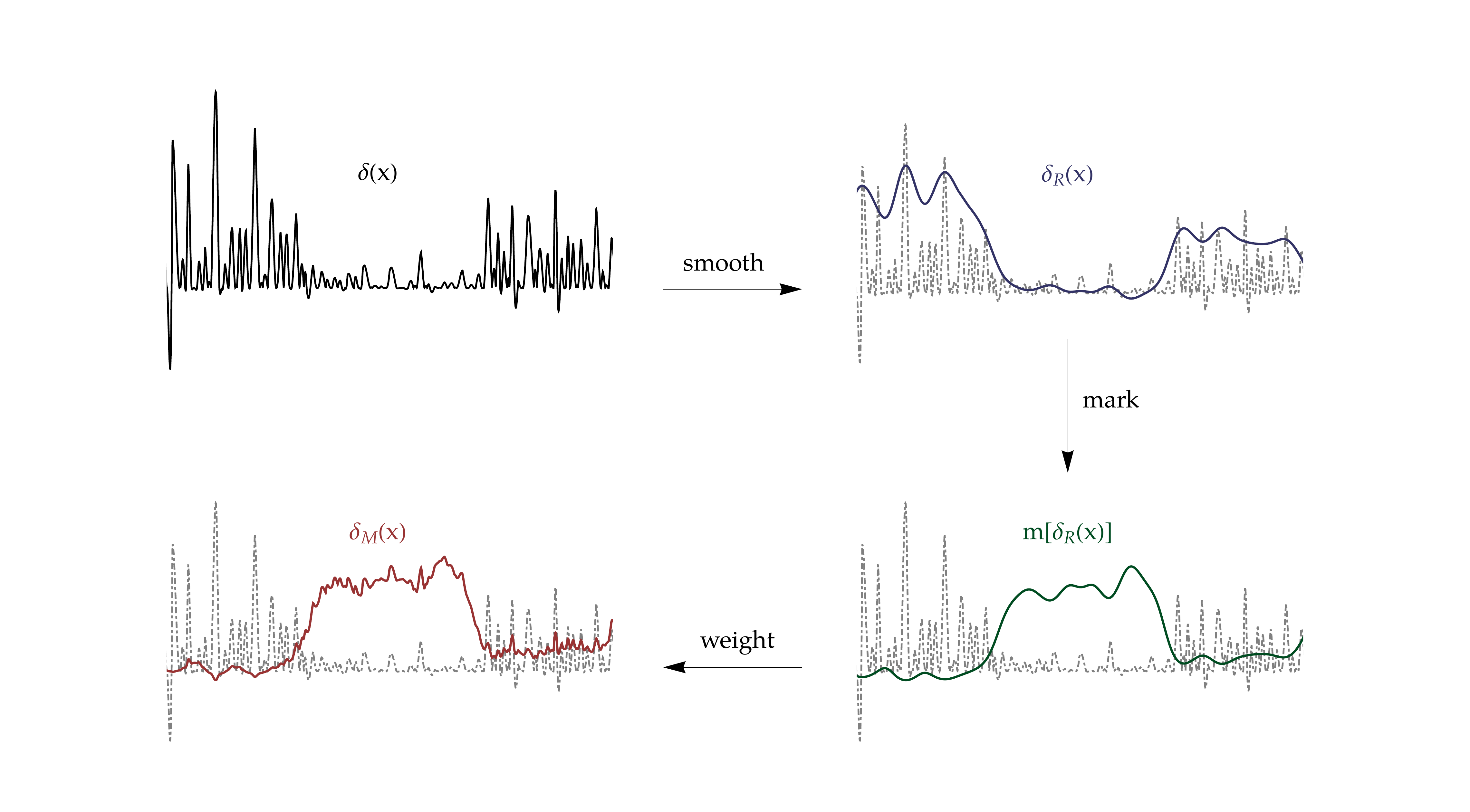}
\caption{1-dimensional cartoon for the construction of the marked density field. }
\label{fig:markedDensity1D}
    \end{center}
\end{figure}

The construction of a marked density field $\delta_M(\vx,t)$ is a three step process, depicted in figure \ref{fig:markedDensity1D}. It goes as follows, 
\begin{enumerate}
    \item One starts with the matter fluctuations field $\delta(\vx,t)$ and convolve it with a window function $W_R(\vx)$ of width $R$ (a Gaussian is used in this work)  to obtain the smoothed density 
    \begin{equation}
        \delta_R(\vx,t) = \int  d^3 y \; W_R(|\vx-\ve y|) \delta(\ve y,t). 
    \end{equation}
    The scale $R$ is chosen from the beginning and it determines the size of the region which is used to define the environmental density of the tracers. In this sense, it is good to have sufficiently large values for $R$, typically comparable to, or a little bit smaller than, the size of cosmic voids. Here we use $R=10 \,h^{-1}\,\text{Mpc}$.
    
    
    \bigskip
    
    \item Construct the mark function $m(\delta_R(\vx))$. We choose a function that decays with the environmental (or smoothed) density, such that low density regions are up-weighted, as required. For example, a simple exponential $m = \exp [ -\alpha (1+\delta_R(\vx)) ]$ can do the job. However it is better to have more free parameters such that we have more liberty to find an optimal mark for a particular MG model. The most used weight is the White-mark\cite{White:2016yhs}
    \begin{equation}\label{MWmark}
        m(\delta(\vx),R) = \left( \frac{1+\delta_*}{1+\delta_* + \delta_R(\vx)}\right)^p,
    \end{equation}
    with $p>0$ and $\delta_*>0$ dimensional parameters. It was find with numerical simulations that the choices $p=10$ and $\delta_*=4$ maximize both the Fisher information and the signal-to-noise.\cite{Valogiannis:2017yxm} However, these parameters are not good for perturbative analysis, so we will use the parameters $p=7$ and $\delta_*=10$, unless otherwise is stated. Other authors working with simulations have used different parameters and mark functions.\cite{Valogiannis:2017yxm,Hernandez-Aguayo:2018yrp,Armijo:2018urs,Alam:2020jdv}

    \bigskip
    
    \item Weight the density field of tracers $1+\delta_X(\vx)$ with the mark function $m(\delta(\vx),R)$ to obtain the  marked density field
    \begin{equation}
    1+\delta_M(\vx,t) =     \big( 1+\delta_X(\vx,t) \big) m[\delta_R(\vx)].
    \end{equation}
    
    The density field can be that of any tracer of the underlying matter density, it can even be the matter density field itself. But we observe tracers, as galaxies for example, in the sky. These are related through a biasing function $F$ to the dark matter as
    \begin{equation}\label{defFx}
        1+\delta_X(\vx) = F_\vx[\delta(\vx), \nabla^2\delta(\vx),...; \vx]
    \end{equation}
    where the stochasticity of this relation is emphasized by writing $\vx$ as an additional argument. The dependence on the non-local curvature operator $\nabla^2\delta(\vx)$ is expected to be small in GR, but a well defined bias expansion within chameleon theories must contain higher-order derivative operators.\cite{Desjacques:2016bnm,Aviles:2018saf}

\end{enumerate}

\section{Marked correlation function}

A marked correlation function (mCF) \cite{Beisbart:2000ja,Gottloeber:2002vm,Sheth:2005ai,Sheth:2005aj,Skibba:2005kb,White:2008ii,White:2016yhs} is defined as the sum of pairs of objects ($i,j$) separated by a distance $r$, and weighted by the ratio of the mark function value to the mean mark $m_i/\bar{m}$ at each point and  divided by the number of pairs $n(r)$,
\begin{equation}\label{mCF}
 \mathcal{M}(r) = \sum_{i,j|r_{ij}=r}\frac{m_i m_j}{n(r)\bar{m}^2},
\end{equation}
so it quantifies the deviation of a pair of objects to have the mean mark $\bar{m}$ subjected to be located at a separation $r$. By construction it can be written as
\begin{equation}
   \mathcal{M}(r) = \frac{1+W(r)}{1 + \xi(r)} 
\end{equation}
where $\xi(r)$ is the standard 2-point correlation function of tracers
\begin{equation}
    1+\xi(r) = \left\langle \big(1+\delta_X(\vx+\vr)\big)  \big(1+\delta_X(\vx)\big) \right\rangle, 
\end{equation}
and $W(r)$ is the 2-point correlation function of marked fields:
\begin{equation}
    1+W(r) = \frac{1}{\bar{m}^2} \left\langle \big(1+\delta_M(\vx+\vr)\big)  \big(1+\delta_M(\vx)\big) \right\rangle. 
\end{equation} 
At large scales both $\xi$ and $W$ are small and one gets   
\begin{equation}
   \mathcal{M}(r) = 1+ W(r)- \xi(r).  
\end{equation}
Now, one may show that $\xi(r) \in W(r)$,\cite{Aviles:2019fli} hence the mark function becomes a 2-point statistic of the clustering of marks with the clustering of objects effectively factorized out. The presence of the mean mark is not necessary to estimate the mCF, but it plays an important role in the analytical renormalization of the theory.

\begin{section}{Perturbative treatment of the marked correlation function} \label{sec:Eulerian}

It is well known that a good modeling of N-point statistics of density fields in configuration space requires the use of Lagrangian Perturbation Theory (LPT),\cite{White:2014gfa} since the spread and degradation of the BAO oscillations is due to large scale bulk flows, long wave-length Lagrangian displacement fields that are well described in the Lagrangian framework, even at the linear order.\cite{Baldauf:2015xfa}   
However the derivation of the marked correlation function within the Lagrangian framework is quite large and technical, so we prefer to omit it in this note;  the interested reader can find it in Ref.~\citenum{Aviles:2019fli}, and later on we will display the main results. However, in this section we develop the Standard Perturbation Theory (SPT) approach to the marked correlation function, which is easier, but yet shows the main ingredients and insights of the theory. 

We start by expanding the mark in a power series of $\delta_R$
\begin{equation}\label{Taylorm}
m(\delta_R; C_i)  =C_0 + C_1 \delta_R + \frac{1}{2} C_2 \delta^2_R + \cdots
\end{equation}
with $C_i =  m^{(i)}[0]$, the $i$-th derivative of $m[\delta_R]$ evaluated at $\delta_R=0$. 
The fewer parameters are needed to model the mark, the better the convergence will be. 
We notice that $C_1<0$  enhances low density regions.

For the sake of simplicity we will assume a local relation between matter and tracer overdensities
\begin{equation}
    1+\delta_X(\vx) = c_0+c_1 \delta(\vx) + \frac{c_2}{2}\delta^2(\vx) + \cdots, 
\end{equation}
with $c_i =\partial F_\vx^{(i)}[\delta_R=0] /\partial \,\delta_R$ the bare bias parameters and $F_\vx$ given by eq.~(\ref{defFx}).
A Taylor expansion is an unnatural and restrictive assumption for biased tracers, but being pragmatic we note that it works at the level of precision required by our simulations.

The mean mark is given by the mark weighted by the tracer density field 
\begin{align} \label{prebarmE}
\bar{m} &= \langle m[\delta_R(\vx)](1+ \delta_X(\vx)) \rangle \nonumber\\
&= \Big \langle  \Big(C_0+C_1 \delta_R(\vx) + \frac{C_2}{2}\delta^2_R(\vx) + \cdots \Big)
\Big(c_0+c_1 \delta(\vx) + \frac{c_2}{2}\delta^2(\vx) + \cdots \Big) \Big \rangle \nonumber\\
&=c_0C_0+ c_1C_1 \sigma^2_R  + \frac{1}{2}c_0C_2 \sigma^2_{RR}  + \frac{1}{2}C_0c_2 \sigma^2 + \cdots,
\end{align}
with zero-lag correlators defined as
\begin{equation}
\sigma^2 \equiv \langle (\delta(0))^2\rangle, \quad \sigma^2_R \equiv \langle \delta_R(0)  \delta(0)\rangle, \quad \sigma^2_{RR} \equiv \langle(\delta_{R}(0))^2\rangle.
\end{equation}
We will use also the correlation and cross-correlation functions
\begin{align}
\xi(r) &\equiv \langle \delta(\vx)\delta(\vx+\vr)\rangle, \label{xiL}\\
\xi_R(r) &\equiv \langle \delta_R(\vx)\delta(\vx+\vr)\rangle, \label{xiR} \\
\xi_{RR}(r) &\equiv \langle\delta_R(\vx)\delta_R(\vx+\vr)\rangle.  \label{xiRR} 
\end{align}
We have to write eq.~(\ref{prebarmE}) in terms of renormalized bias parameters,\cite{Matsubara:2008wx}
\begin{equation}\label{bnEdef}
 b_n^E = \int \frac{d\lambda}{2\pi} e^{-\lambda^2 \sigma^2/2} \tilde{F_\vx}(\lambda) (i\lambda)^n, 
\end{equation}
where $\tilde{F}_\vx(\lambda)$ is the Fourier transform of $F_\vx$ with spectral parameter $\lambda$ (dual to $\delta$). 
We are using  
the label ``$E$'' to distinguish Eulerian from Lagrangian biases. Equation (\ref{bnEdef}) leads to the 
relation between bare bias $c_n$ and renormalized bias $b_n^E$ parameters:\cite{Aviles:2018thp,Eggemeier:2018qae} 
\begin{equation}
  b_n^E=\sum_{k=0}^{\infty} \frac{\sigma^{2k} }{2^k k!} c_{n+2k}.  
\end{equation}
Analogously, we introduce the ``resummed'' expansion parameters $B_n$ as
\begin{equation}\label{Bndef}
B_n=\frac{B_n^*}{B_0^*} \qquad \text{with} \qquad B_n^* = \int \frac{d\Lambda}{2\pi} e^{-\Lambda^2 \sigma^2_{RR}/2} \tilde{m}(\Lambda) (i\Lambda)^n,
\end{equation}
where $\tilde{m}(\Lambda)$ is the Fourier transform of $m(\delta_R)$, and $\Lambda$ is a spectral parameter, dual to $\delta_R$. One easily 
finds\cite{Aviles:2019fli} the following relation between the expansion parameters $B_n$ and the Taylor coefficients of the mark function $C_n$
\begin{equation}\label{BnRenorm}
B_n(C_n,\sigma_{RR}^2) = \frac{\sum_{k=0}^{\infty} C_{n+2k} \sigma^{2k}_{RR}/(2^k k!)}{\sum_{k=0}^{\infty}  C_{2k} \sigma^{2k}_{RR}/(2^k k!)}.
\end{equation}
Inserting the $b_n^E$ and $B_n$ parameters in eq.~(\ref{prebarmE}) we get
\begin{equation} \label{barmE}
 \bar{m} = B_0^* \big[ 1 + b_1^E B_1 \sigma^2_R + \cdots \big].
\end{equation}

We compute now 
\begin{align}
&\bar{m}^2(1+W(r)) = \langle m[\delta_R(\vx_1)]\big(1+\delta_X(\vx_1)\big)   m[\delta_R(\vx_2)]\big(1+\delta_X(\vx_2)\big) \rangle 
\nonumber \\
&= \int \frac{d\lambda_1 d\lambda_2 d\Lambda_1  d\Lambda_2}{(2 \pi)^4} \langle e^{ i(\lambda_1\delta_1 +\Lambda_1\delta_{R,1} + \lambda_2\delta_2 +\Lambda_2\delta_{R,2})} \rangle \tilde{F}_\vx(\lambda_1)\tilde{F}_\vx(\lambda_2)
\tilde{m}(\Lambda_1)\tilde{m}(\Lambda_2) \nonumber\\
&= \int \frac{d\lambda_1 d\lambda_2 d\Lambda_1  d\Lambda_2}{(2 \pi)^4} \tilde{F}_\vx(\lambda_1)\tilde{F}_\vx(\lambda_2)
\tilde{m}(\Lambda_1)\tilde{m}(\Lambda_2) 
 e^{ -\frac{1}{2}(\lambda_1^2 + \lambda_2^2)\sigma^2 -\frac{1}{2}(\Lambda_1^2 + \Lambda_2^2)\sigma^2_{RR} } \nonumber\\
 &\quad \times \Big[1 - (\lambda_1\Lambda_1 + \lambda_2\Lambda_2 )\sigma^2_R   - \lambda_1\lambda_2 \xi(r) -\Lambda_1\Lambda_2 \xi_{RR}(r) - 
 (\lambda_1\Lambda_2+\lambda_2\Lambda_1)\xi_R(r) + \cdots \Big] \nonumber\\
&= \bar{m}^2\Big[ 1+(b_1^E)^2 \xi(r)+B_1^2 \xi_{RR}(r) + 2 b_1^E B_1 \xi_{R}(r) +\cdots \Big],
\end{align}

\begin{figure}
    \begin{center}
\includegraphics[width=3in]{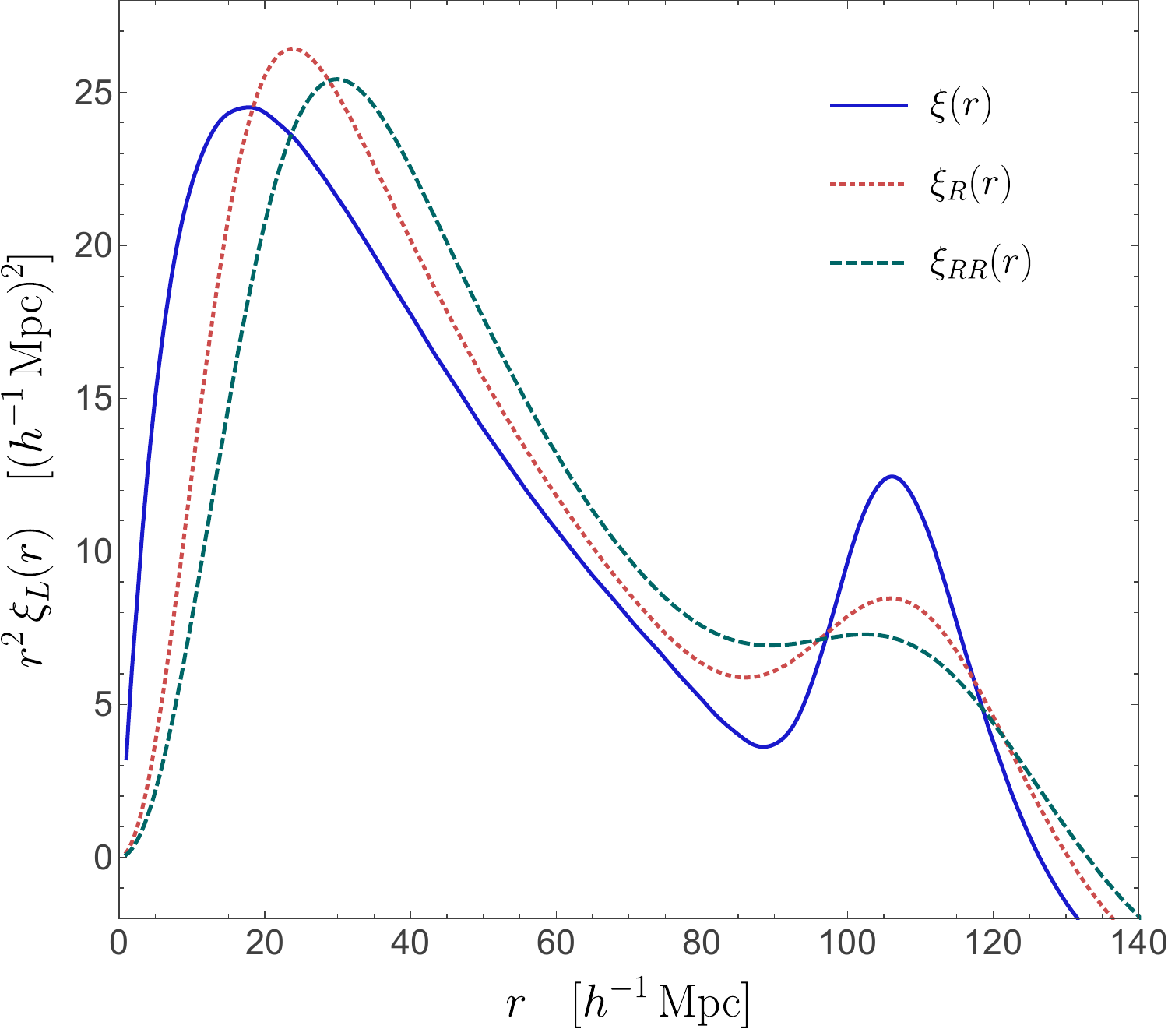}
\caption{Smoothed correlation functions defined in defined in eqs.~(\ref{xiL})-(\ref{xiRR}) and appearing in the Eulerian marked correlation function [Eq.~\eqref{mCFE}]. We use a Gaussian damping kernel $W_R$ of width $R=10\,h^{-1} \text{Mpc}$ to smooth the density fields.}
\label{fig:xiRR}
    \end{center}
\end{figure}

where the ellipsis denotes second order terms in $\xi_{{},R,RR}(r)$. In the second equality, we have shifted to Fourier space ($\delta \rightarrow \lambda$, $\delta_R \rightarrow \Lambda$)  and  in the third equality we used the cumulant expansion theorem, 
\begin{equation} \label{CET}
 \langle e^{iX} \rangle = \exp\left[ \sum_{N=1}^\infty \frac{i^N}{N!} \langle X^N \rangle_c \right],
\end{equation}
and expanded out of the exponential all terms but those containing
$\sigma^2$ and $\sigma^2_{RR}$, such that we can use Eqs.~(\ref{bnEdef}) and (\ref{Bndef}) to get biases from spectral 
parameters, as we did in the last equality.
The mCF becomes
\begin{equation} \label{mCFE}
\mathcal{M}^E(r) =  \frac{1+W(r)}{1+\xi_X(r)} = \frac{1+(b_1^E)^2 \xi(r)+B_1^2 \xi_{RR}(r) + 2 b_1^E B_1 \xi_{R}(r) + \cdots}{1+(b_1^E)^2 \xi(r) + \cdots}.
\end{equation}
In figure \ref{fig:xiRR} we show the correlation functions defined in eqs.~(\ref{xiL})-(\ref{xiRR}) and used to construct the above mCF. We use a Gaussian smoothing kernel $W_R$ with width $R=10\,h^{-1} \text{Mpc}$.

We  notice that the zero-lag correlators $\sigma^2$ and $\sigma^2_{RR}$ do not appear in 
the mCF as is guaranteed because we are using renormalized $b$ and $B$ parameters. 
Meanwhile, the cross-covariance $\sigma^2_R$ is canceled out by the mean mark squared appearing 
in the definition of the mCF in eq.~(\ref{mCF}).  We remark that a  process of renormalization of the Taylor expansion
coefficients $C$ is not strictly necessary because the scale $R$ is physical, and chosen from the beginning by the observer to mark the tracers, so there is no impediment for $\sigma_{RR}^2$ to appear in the final expressions. However, the use of parameters $B$ instead of the Taylor coefficients leads to simpler final equations. 
Furthermore, in real applications, we have to use a mark computed by the number density of galaxies. Another advantage of using the renormalized $B$ parameters is that 
the effect of this reassignment can be included simply by re-scaling the $B$ parameters. For the application to dark matter halos and galaxies, one should treat them as free parameters and fit them with simulations. 

 \end{section}

\begin{section}{Lagrangian space}

In Lagrangian space one considers regions of space, at an initial early time with spatial coordinates $\vq$. We relate matter and tracers overdensities, in an analogous way to eq.~(\ref{defFx}), by
\begin{equation}\label{defF}
 1+\delta_X(\vq) = F[\delta(\vq)],
\end{equation}
with the difference that the overdensity fields are taken at an initial early time.\footnote{This initial time is taken early enough that all scales of interest are still in the linear regime and one can safely take $\delta(\vq)\ll 1$, but late enough that the background universe is already in the Einstein-de Sitter phase.}
Assuming number conservation of tracers, $(1+\delta_X(\vx))d^3x = (1+\delta_X(\vq))d^3q$, one can relate the function $F_\vx$, introduced in eq.~(\ref{defFx}), and $F$  
through
\begin{equation}\label{FxToF}
 F_\vx[\delta(\vx)] = \int \Dk{k} \int d^3 q \, e^{i\vk\cdot(\vx-\vq)} \int\frac{d\lambda}{2\pi} \tilde{F}(\lambda) e^{i\lambda\delta(\vq) - i \vk\cdot \mathbf{\Psi(\vq,t)}},
\end{equation}
with $\mathbf{\Psi}(\vq)$ the Lagrangian displacement vector, that maps Lagrangian coordinates $\vq$ to Eulerian coordinates $\vx$ as $\vx(\vq,t) = \vq + \mathbf{\Psi}(\vq,t)$.
Equivalently to eq.~(\ref{bnEdef}), we introduce the renormalized Lagrangian local biases \cite{Matsubara:2008wx} with
\begin{equation}\label{bndef}
 b_n = \int \frac{d\lambda}{2\pi} e^{-\lambda^2 \sigma^2/2} \tilde{F}(\lambda) (i\lambda)^n. 
\end{equation}
The next step is to evolve initially biased tracers with overdensity $\delta_X(\vq)$ 
using Convolution Lagrangian perturbation theory\cite{Carlson:2012bu,Vlah:2015sea} 
and thereafter assign
them a mark $m[\delta_R(\vx)]$ at the moment of observation.
The use of the Lagrangian approach has some advantages 
with respect to the Eulerian. In the first place it is 
well known that the two-point correlation function is poorly 
modeled within the Eulerian approach, particularly at the BAO peak position; 
second, the (renormalized) Lagrangian bias parameters are obtained through 
the peak background split prescription \cite{Kaiser:1984sw,Mo:1996cn,Sheth:1999mn,Schmidt:2012ys}, and hence are physically appealing. But the price to pay is that the mCF is cumbersome to compute, since it becomes a double Gaussian convolution\cite{Aviles:2019fli}: 
\begin{align} \label{mCFCLPT2}
1+W(r) &= 
\int  \frac{d^3 q \, e^{-\frac{1}{2}(\vr-\vq)^T\mathbf{A}{-1}(\vr-\vq)} }{(2\pi)^{3/2}|\mathbf{A}|^{1/2}}
\int  \frac{d^3 Q \,e^{-\frac{1}{2}(\ve R-\ve Q)^T\mathbf{C}^{-1}(\ve R-\ve Q)}  }{(2\pi)^{3/2}|\mathbf{C}|^{1/2}}  \\
&\quad \qquad \times \Big(1+ \mathcal{I}^0(\vr,\vq,\ve R, \ve Q;b_i,B_i) \Big),
\end{align}
with $\mathcal{I}^0$ a function, whose expression is quite large and not very illuminating, and contains the information of the mark (through the expansion parameters $B_i$) and of the tracers (through the bias parameters $b_i$). It can be found in eq.~(B.8) of Ref.~\citenum{Aviles:2019fli}. The matrices $\ve A$ and $\ve C$ are given by
\begin{equation}\label{AijL}
 A_{ij}(\vq) = 2 \int \Dk{p} \big( 1 - e^{i\vp\cdot \vq} \big) \frac{p_ip_j}{p^4}P_L(p),
\end{equation} 
and
\begin{equation}
 C_{ij}(\vq) = - \frac{1}{4}A_{ij}(\vq) + \frac{1}{6\pi^2}\int_0^\infty P_L(p) dp \,\,\delta_{ij},
\end{equation}
with $P_L(p)$ the linear matter power spectrum.  
We notice that if $1+ \mathcal{I}$ is not a function of $\ve Q$, it
can be pulled out of the $\ve Q$ integral and due to that $C_{ij}$ depends only on $\vq$, 
the integration over $\ve Q$ gives 1. This is the case of the ``standard'' correlation function in Convolution Lagrangian Perturbation Theory
(CLPT),\cite{Carlson:2012bu} reducing the double Gaussian convolution in eq.~(\ref{mCFCLPT2}), to a single three-dimensional convolution.  
\begin{figure}
	\begin{center}
	\includegraphics[width=3 in]{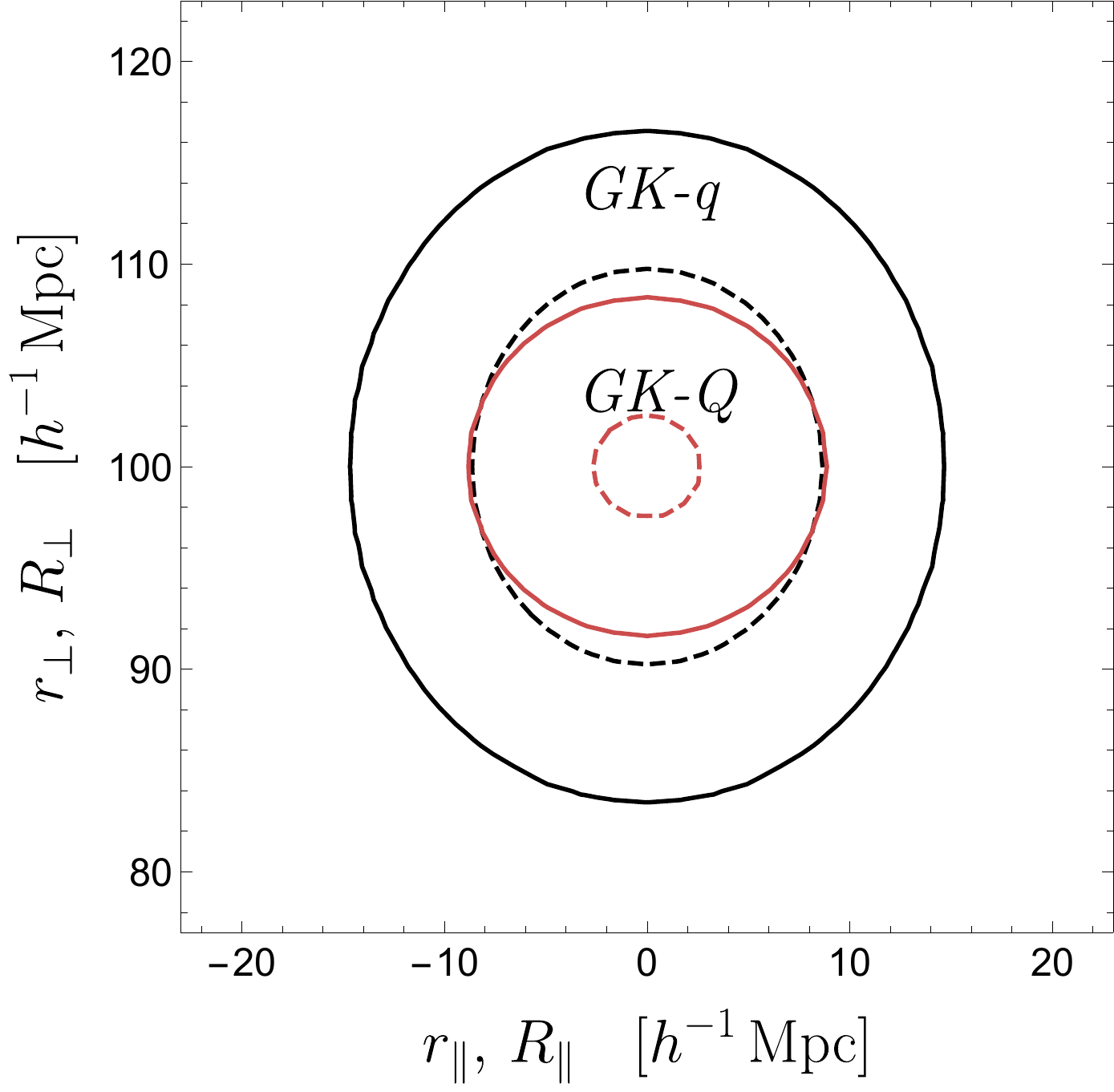}	
	\caption{Gaussian kernels in eq.~(\ref{mCFCLPT2}). {\it GK}-$q$ is the kernel of the $q$-integral (black curves) and 
	{\it GK}-$Q$ the kernel of the $Q$-integral (red curves). Dashed and solid curves show the regions that enclose the  68\% and 95\% of the volume
	respectively. A similar plot can be found in Ref.~\citenum{Tassev:2013rta}. Figure adapted from Ref.~\citenum{Aviles:2019fli}.
	\label{fig:GaussianKernels}}
	\end{center}
\end{figure}

In figure~\ref{fig:GaussianKernels} we show level plots for the Gaussian kernels appearing in eq.~(\ref{mCFCLPT2}). 
{\it GK}-$q$ is the kernel of the $q$-integral and 
{\it GK}-$Q$ the kernel of the $Q$-integral. These are shown as a function of $r_\parallel$ ($R_\parallel$) and $r_\perp$ ($R_\perp$), 
the components of $\vr$ ($\ve R$) parallel and perpendicular
to the Lagrangian coordinate $\vq$ ($\ve Q$) with $Q=q=100\, h^{-1} \,\text{Mpc}$ fixed. To have a sense of the meaning of this plot, consider that the kernel {\it GK}-$q$ is the
probability distribution of finding two dark matter particles separated by a distance $\vr$ at the redshift of evaluation ($z=0.5$ in the plot), given that they were separated by a distance $\vq$ at the early initial time.$\!$\cite{Tassev:2013rta}
We notice that both kernels have their maximum value at
$Q=q=100\, h^{-1} \,\text{Mpc}$, but the {\it GK}-$Q$ is more sharply peaked because $|\mathbf{C}|<|\mathbf{A}_L|$. 
This observation suggests to approximate 
\begin{equation}
\frac{  e^{-\frac{1}{2}(\ve R - \ve Q)^T \mathbf{C}^{-1}(\ve R - \ve Q) }  }{(2\pi)^{3/2}|\mathbf{C}|^{1/2}} \approx \dD(\ve R-\ve Q).
\end{equation}
By doing so, one arrives at
\begin{equation}\label{1pWM}
1+ W^\text{W16}(r) = \int \frac{d^3 q}{(2\pi)^{3/2}|A_L|^{1/2}} e^{-\frac{1}{2}(\vr-\vq)^T\mathbf{A}^{-1}_L(\vr-\vq)}
\Big(1+ J^{0}(\vq,\vr;b_i,B_i) \Big)
\end{equation}
which is the generalization of the Zeldovich approximation (ZA) result obtained in Ref.~\citenum{White:2016yhs}. The expression for 
$J^{0}(\vq,\vr;b_i,B_i)$ is given by eq.~(4.18) of Ref.~\citenum{Aviles:2019fli}.

\begin{figure}
	\begin{center}
	\includegraphics[width=5 in]{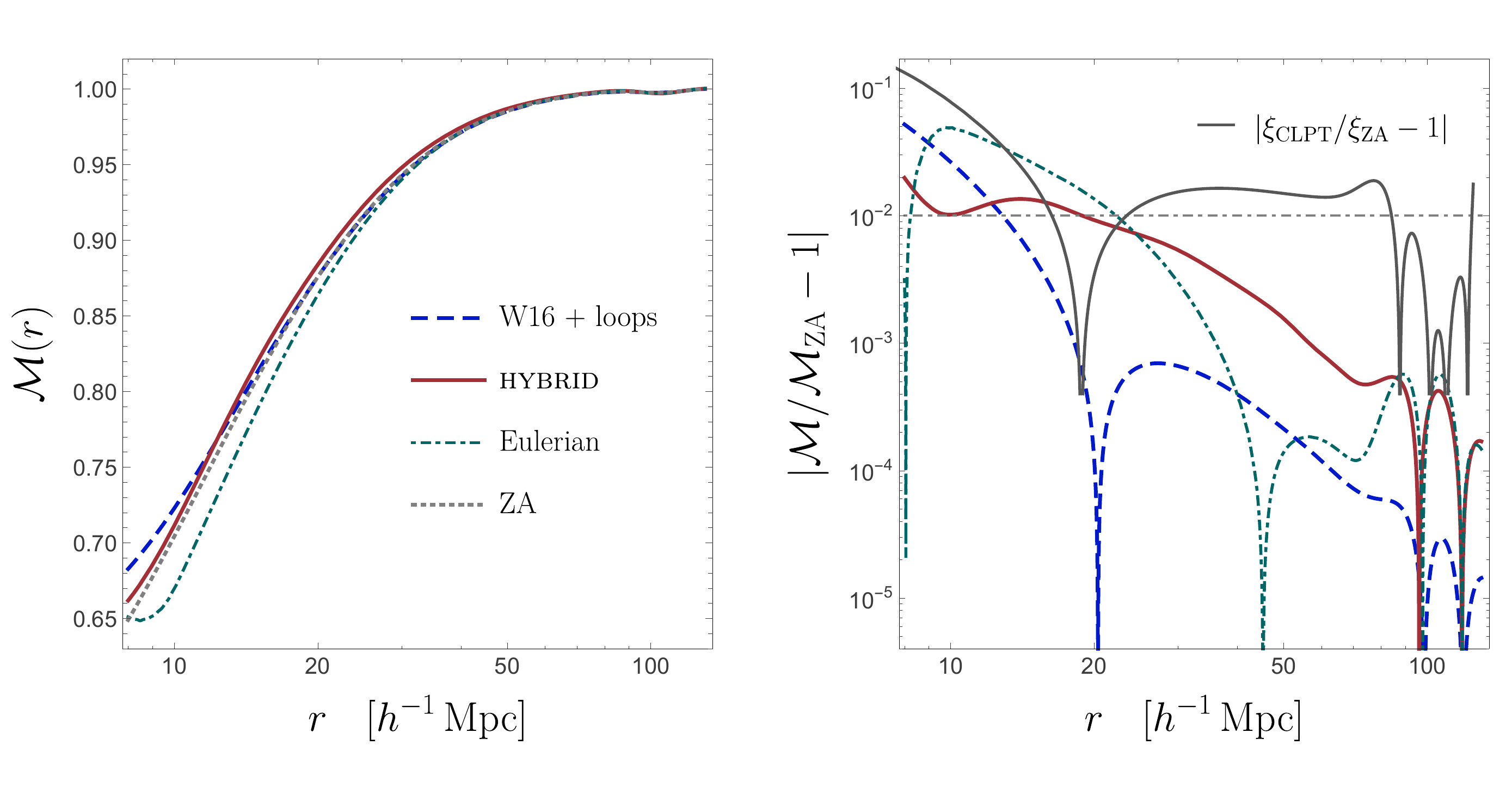}	
	\caption{Marked correlation function modeling comparison. 
	The left panel shows the mCF with parameters 
	$b_1=1.2$, $b_2=0.2$, $B_1=-0.7$, $B_2=B_1^2$, for the different methods:
	solid red curve is the \textsc{hybrid} model presented in Ref.~\citenum{Aviles:2019fli}; dashed blue the method of  Eq.~(\ref{1pWM}), which is the method introduced in Ref.~\citenum{White:2016yhs}, plus 1-loop nonlinear contributions; in dotted gray, the ZA; and in dot-dashed green, the Eulerian linear model of eq.~(\ref{mCFE}). The right panel shows the relative differences with respect to the ZA.  
	The gray dot-dashed horizontal line denotes the $1\%$ differences. The solid black curve in the right panel shows the relative differences between the CLPT 1-loop and ZA standard correlation functions. Figure adapted from Ref.~\citenum{Aviles:2019fli}.
	\label{fig:CompareModels}}
	\end{center}
\end{figure}

Figure \ref{fig:CompareModels} shows the mCF using different analytical methods for the MG Hu-Sawicki F5 model: solid red curve is the one presented above and developed in Ref.~\citenum{Aviles:2019fli}---called \textsc{hybrid} from now on, because it evolves the tracers using LPT but marks the tracers using the Eulerian smoothed matter overdensity fields at the moment of observation. The dashed blue is the method of  eq.~(\ref{1pWM}), which is that introduced in Ref.~\citenum{White:2016yhs} with the addition of leading non-linear (loop) corrections; in dotted gray, the Zeldovich Approximation, 
which in this work means the linear model of Ref.~\citenum{White:2016yhs}; and, in dotdashed green, the Eulerian linear model of eq.~(\ref{mCFE}).
We are using Lagrangian local biases $b_1=1.2$ and $b_2=0.2$, and mark parameters $B_1=-0.7$, $B_2=B_1^2$. In the right panel we show the relative difference between the models and the ZA. The  agreement between them is very good, even for the linear Eulerian theory, below 1\% for scales above $\sim 20 \,h^{-1}\,\text{Mpc}$. The relative differences between the CLPT standard correlation function with and without loop contributions ($\xi_\text{CLPT}$ and $\xi_\text{ZA}$, respectively) is shown in the right panel (black curve); by comparing it with the dashed blue curve (which shows the relative differences of the  W16+1-loop and ZA mCFs) we confirm that mCFs which efficiently enhance low density regions in the sky are indeed more linear than the standard correlation function.

\end{section}

\begin{section}{Comparison to simulations}

We use the \texttt{Elephant} suite of simulations run with the \texttt{ECOSMOG} code. \cite{2012JCAP...01..051L} This suite contains five realisations of the initial conditions and for each realisation we have one simulation of $\Lambda$CDM (GR) together with three simulations of the Hu-Sawicki $f(R)$ model with parameters $f_{R0} = -10^{-4}$ (F4), $f_{R0} = -10^{-5}$ (F5) and $f_{R0} = -10^{-6}$ (F6). It also contains galaxy mock catalogs that were made with a Halo Occupation Distribution (HOD) method. The HOD parameters for the $\Lambda$CDM model are the best-fit parameter values from the CMASS data.\cite{2013MNRAS.428.1036M} For the $f(R)$ models, we tune the HOD parameters so that we reproduce the correlation function in the $\Lambda$CDM model. The simulations were ran in a box of size $L=1024\, h^{-1} \,\text{Mpc}$ with $N = 1024^3$ particles and the cosmological parameters used to make the initial conditions were $\Omega_b = 0.046$, $\Omega_\Lambda = 0.719$, $\Omega_m = 0.281$, $h = 0.697$, $\sigma_8 = 0.82$ and $n_s = 0.971$, and for our analysis we choose snapshots at redshift $z=0.5$.
To compute the mark for each of our tracers (dark matter particles, halos or mock galaxies) we binned the particles/halos/mock galaxies to a grid with gridsize of $20\, h^{-1} \,\text{Mpc}$ (corresponding to a $N = 52^3$ grid) using a Nearest Grid Point assignment scheme to get an estimate for the density for which the mark depends on.

Both, the standard and weighted 2-point correlation functions were computed using the Correlation Utilities and Two-point Estimates (\texttt{CUTE})\cite{2012arXiv1210.1833A} code.\footnote{https://github.com/damonge/CUTE} From this the marked correlation function follows simply as $\mathcal{M}(r) = (1+W(r))/(1+\xi(r))$. 

We consider the White-mark with $\rho_*=10$, $p=7$, corresponding 
to coefficients $C_0=1$, $C_1=-0.64$, $C_2=0.46$ in the 
Taylor expansion of the mark function in eq.~(\ref{Taylorm}). In our analytical models
we smooth the matter fields that assign the mark with a top-hat filter $W_R$ of radius $R=10\, h^{-1} \,\text{Mpc}$. 

\begin{figure}
    \begin{center}
	\includegraphics[width=3.0 in]{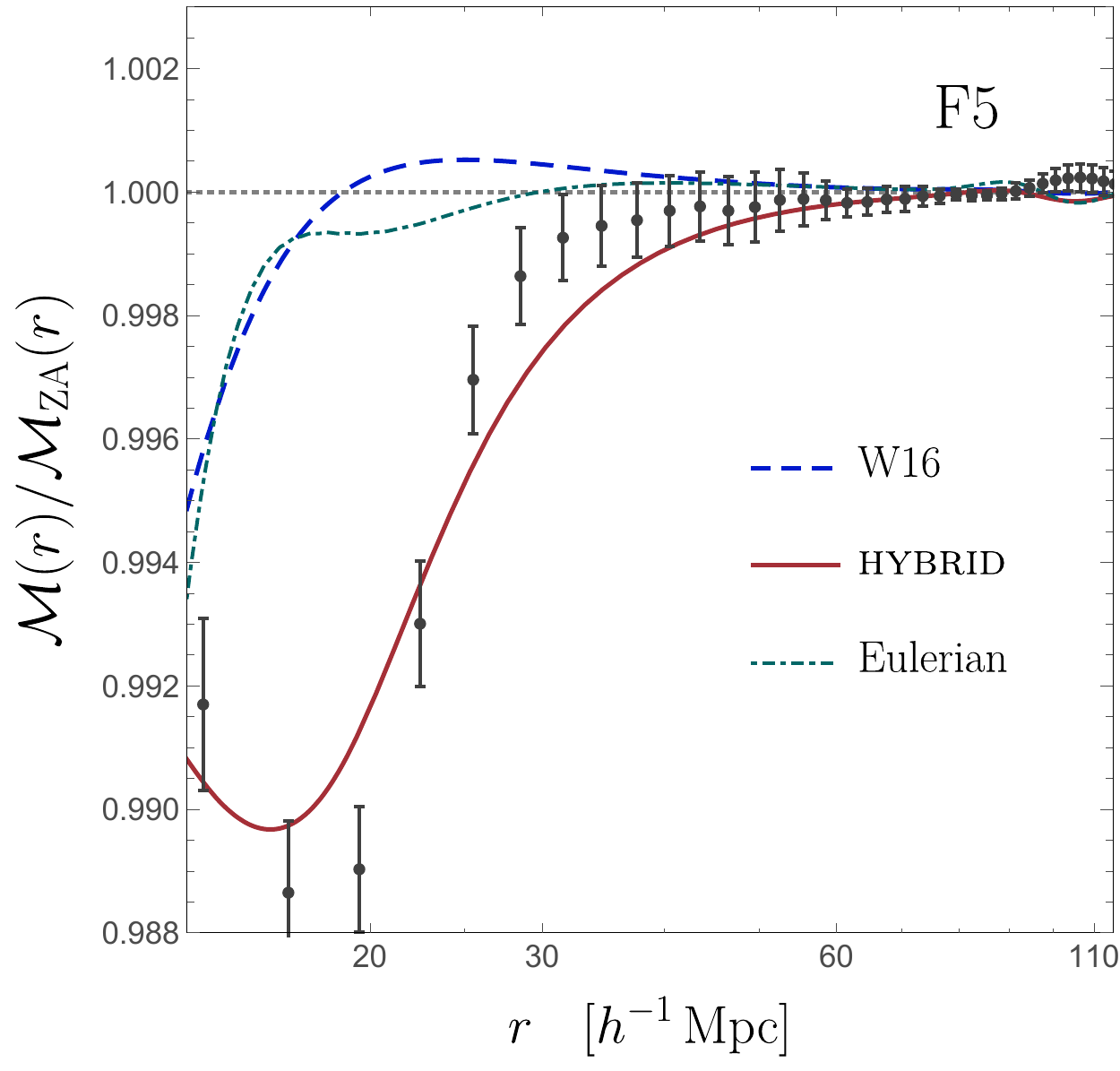}		
	\caption{Ratios of different mCFs analytical models to the ZA mCF for 
	the F5 gravity (we obtain similar results for F4, F6 and GR). We show the W16 model plus 1-loop corrections (dashed blue); 
	the \textsc{hybrid} model (solid red); and the linear Eulerian model of eq.~(\ref{mCFE}) (dot-dashed green).  Figure adapted from Ref.~\citenum{Aviles:2019fli}.
	\label{fig:mattermCFsratios}}
	\end{center}
\end{figure}

We first confront different analytical methods against matter data, since in this case the mCF does not suffer from biasing-marking degeneracies and we can observe more neatly the effects of the weights. 
The expansion parameters $B$ are obtained from eq.~(\ref{BnRenorm}): $B_1^\text{GR}=-0.6495 $, $B_1^\text{F6}=-0.6497$, 
$B_1^\text{F5}=-0.6505$, $B_1^\text{F4}=-0.6522$, $B_2^\text{GR}= 0.4628$, $B_2^\text{F6}=0.4502$, $B_2^\text{F5}=0.4495$, $B_2^\text{F4}=0.4480$; which are numerically very close to $C_1$ and $C_2$ values because $C_0=1$ and the variances $\sigma^2_{RR}$ are small.

In figure~\ref{fig:mattermCFsratios} we show the ratios of the different analytical methods to the ZA model. We do it for the Hu-Sawicki F5 gravity model, and we have checked also the gravity theories GR, F4 and F6, and obtained similar results. The dashed blue curve shows the analytical result with the W16 model plus 1-loop corrections; 
solid red, the \textsc{hybrid} method; and dot-dashed green, the linear Eulerian model of eq.~(\ref{mCFE}).  
The differences between the analytical methods are apparent but small, being lesser than the $1\%$. 
At scales $r>40\, h^{-1} \,\text{Mpc}$, all analytic models are indistinguishable and they are within the errors of the simulation data. 
At smaller scales, the method \textsc{hybrid} 
outperforms the other perturbative approaches and captures reasonably the trend of the data all the way up to the smoothing scale 
$R=10 \,h^{-1}\, \text{Mpc}$.

\begin{figure}
    \begin{center}
\includegraphics[width=2.4in]{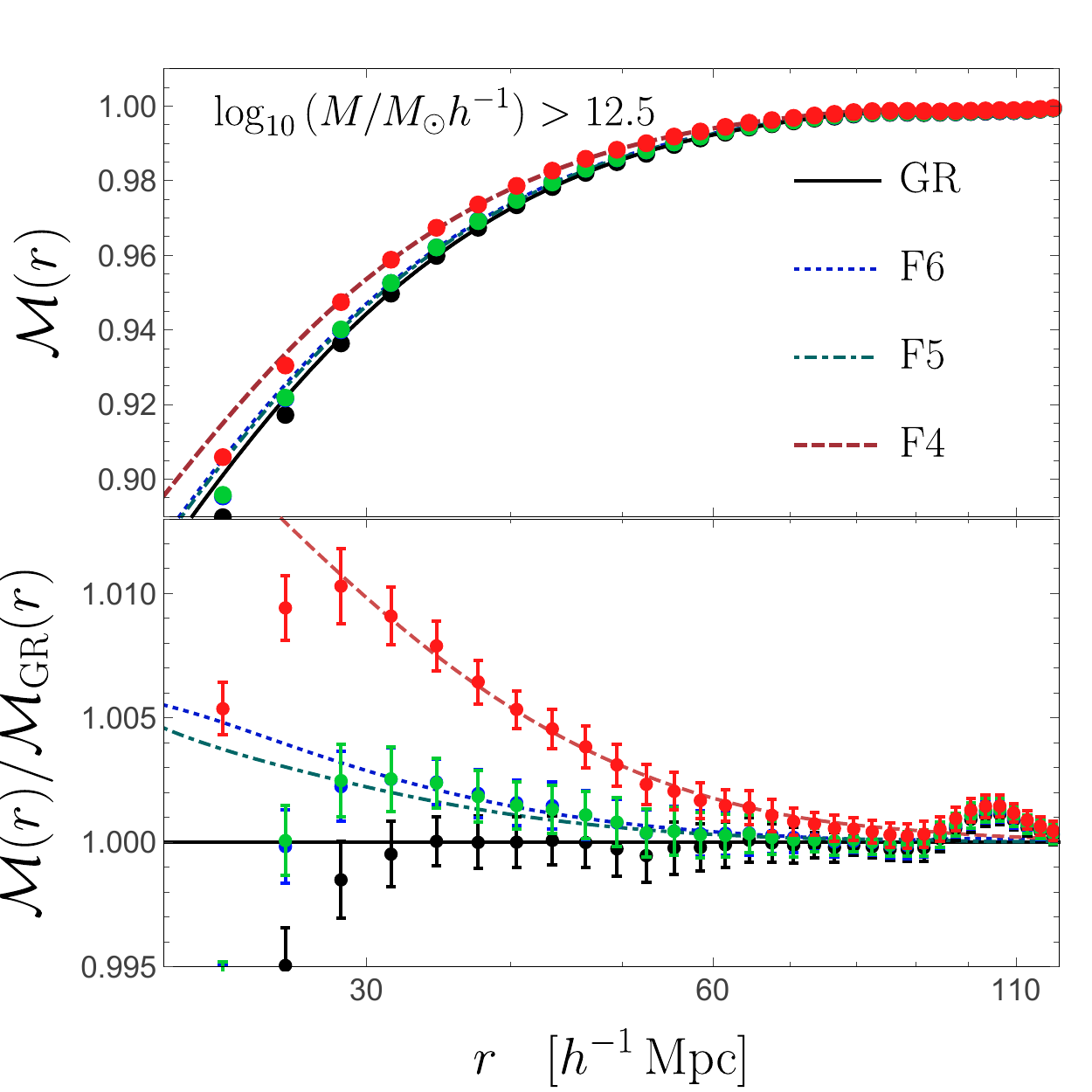}
\includegraphics[width=2.4in]{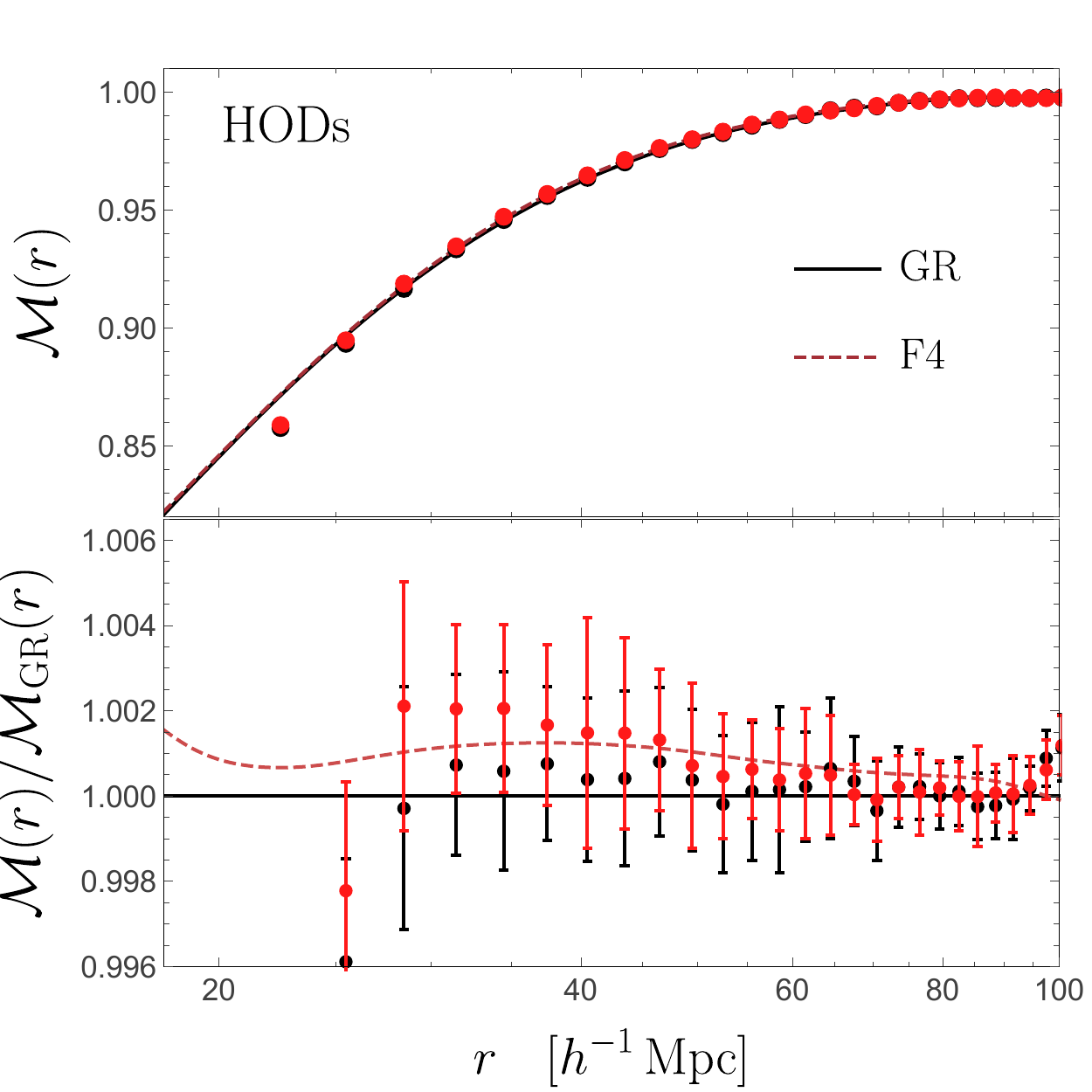}
\caption{Halos and galaxies mCFs. Left panel is for halo masses $12.5 < \log_{10} [M/(M_\odot h^{-1})]<15$ and gravitational models F4, F5, F6 and GR. Right panel is for HOD and shows only the F4 and GR models. The linear local biases are fitted by using the ZA correlation function.  Figure adapted from Ref.~\citenum{Aviles:2019fli}.}
\label{fig:comparingHaloes}
    \end{center}
\end{figure}

Finally, in figure \ref{fig:comparingHaloes} we use the \textsc{hybrid} model to fit the data for halos and galaxies (HOD), finding very good agreement between simulations and theory for $r>30\,h^{-1}\,\text{Mpc}$. In the left panel we show that the MG models are differentiated by the data and the analytical curves in the interval $30<r<50\,h^{-1}\,\text{Mpc}$, with the exception of the F5 and F6 models, which become indistinguishable within the error bars. However, we notice that the HOD mCFs for the different MG models do not look differently enough to be distinguished by our data. We think that the use of different parameters $\delta_*$ and $p$ or different mark functions can help to discern between different gravitational patterns; see, for example, Ref.~\citenum{Alam:2020jdv}.

\end{section}

\begin{section}{Marked power spectrum}
Recently, it was discovered that the marked statistics can be very useful for the estimation of cosmological parameters, particularly the mass scale of the primordial neutrinos.\cite{Massara:2020pli} The reason is that given the large velocity dispersion of neutrinos, they do not contribute to the small scale clustering (see, e.g. Ref.~\citenum{Lesgourgues:2006nd}). But instead, they free-stream over large distances, allowing them to populate the cosmic voids considerably. On the other hand, the influence of neutrinos on high density regions, for example halos, is quite marginal because their contribution to the total matter density is very small. Thus,  statistics  that  up-weight  low  density  regions  can be very  beneficial for parameter estimation. In fact, the Fisher formalism results are quite impressive: the marked power spectrum of the matter field outperforms the constraining power of the standard power spectrum by at least a factor of 2 on all cosmological parameters, and in particular of up to 80 times for the sum of the neutrino masses, when a combination of two different mark functions is used.

The analytical description of the marked power spectrum has been studied in detail in a couple of recent works. First, for matter in real space\cite{Philcox:2020fqx} and later for tracers in redshift space.\cite{Philcox:2020srd} The main complication of such studies is that the mark parameters optimal for improving the cosmological parameter estimation are not able for a perturbative treatment, and the linear theory fails at all scales. The large-scale theory contains non-negligible contributions from all perturbative orders; in Ref.~\citenum{Philcox:2020srd} a reorganization of the theory that contains all terms relevant on large-scales is proposed, with an explicit form at one-loop and the following structure at infinite-loop:
\begin{align*}
 \mathcal{M}(k,\mu) &= \big[C_0 - C_1 W_R(k) \big] \big[ (\tilde{a}_0 + \tilde{a}_2 \mu^2) P_L(k) + \tilde{b}_0 \big] + \text{EFT} + \text{shot noise}  
\end{align*}
with $\tilde{a}_0$, $\tilde{a}_2$ and $\tilde{b}_0$ free parameters of the theory. EFT refers to the standard 1-loop corrections to the linear power spectrum and the effective field theory terms, while the shot noise includes the stochastic contributions.

\end{section}

\begin{section}{Summary}

We have reviewed the analytical study of marked statistics that up-weight low density regions in the Universe.\cite{White:2016yhs,Aviles:2019fli,Philcox:2020srd,Alam:2020jdv} The idea behind marked statistics is to assign a value (the mark) to each entity in a catalogue and perform statistics over the resulting weighted objects. 
An efficient way to suppress non-linearities is by choosing a mark that gives more weight to objects that reside in low 
density regions, which become less important as one consider higher density regions. This marking process can be particularly useful for testing general relativity with cosmological probes, since modified gravity models that provide cosmic acceleration, often rely on screening mechanisms to hide its range and strength on high density environments, and the screenings switch off in regions that are more depleted of matter, like cosmic voids. Therefore, the effects of an hypothetical MG are expected to be more pronounced in low density regions. It is then natural to search for such marked statistics, as proposed in Ref.~\citenum{White:2016yhs}. Further, these statistics can potentially improve the constraining power of all cosmological parameters, in particular the mass of the primordial neutrinos, as it was shown in Ref.~\citenum{Massara:2020pli}.

\end{section}

\section*{Acknowledgments}

I want to thank Alfredo Macias and Carlos Herdeiro for inviting me to give a talk in the BS1 parallel session ``Scalar fields in Cosmology'' of the Sixteenth Marcel Grossmann Meeting -- MG16, July 5-10 2021.
I also want to thank Jorge L. ~ Cervantes-Cota, Baojiu Li, Kazuya Koyama, Hans Winther, Elisa Massara, Oliver Philcox and Martin White for their collaboration in developing the material that led to this work.
I acknowledge support by CONACyT project 283151 and CONACyT Ciencia de Frontera grant No.~102958. 
I would like to thank the DiRAC Data Centric system at Durham University (\url{www.dirac.ac.uk}) for computational facilities. 

\appendix{Hu-Sawicki MG model}\label{app:HS}

The Hu-Sawicki model \cite{Hu:2007nk} is a particular realisation of $f(R)$ gravity that is able to evade the strong constraints coming from local test of gravity and still give rise to interesting observable signatures on cosmological scales. $f(R)$ theories consist on replace the Einstein-Hilbert Lagrangian density $\sqrt{-g} R$ by a general function of the Ricci scalar $\sqrt{-g} \big(R +f(R)\big)$; see Ref.~\citenum{Koyama:2015vza} for a review. 
The Hu-Sawicki model is defined by the function
\begin{equation}\label{fRHS}
  f(R) = - M^2 \frac{c_1 (R/M^2)^n}{c_2 (R/M^2)^n + 1}  
\end{equation}
where the energy scale is chosen to be $M^2 = \Omega_m H^2_0 $. In this parametrized model, at high curvature ($R\ll M^2$)
the function $f(R)$ approaches a constant, recovering GR with cosmological constant, while at low curvature it goes
to zero, recovering GR; the manner these two behaviors are interpolated is dictated by the free parameters. Given
that $d^2 f(R)/dR^2 > 0$ for $R > M^2$,
the solutions are stable and the scalar tensor gravity description is possible.\cite{Magnano:1993bd} In order to be have a similar 
background evolution than the $\Lambda$CDM model it is also necessary that $c_1/c_2 = 6\Omega_\Lambda/\Omega_m$, thus leaving two
parameters to fix the model. One choose these parameters to be $n$ and $f_{R0}\equiv df(R)/dR |_{R=R_0}$, with $R_0=3H_0^2(\Omega_m + 4 \Omega_\Lambda)$ the  Ricci scalar of the background cosmological metric evaluated  nowadays.

In this work we have considered a fixed value $n=1$, and hence the model has only one free parameter $f_{R0}$. Taking this parameter to zero we recover GR. The three choices for the parameters we are considering in this paper (F4, F5 and F6) are such that they lie in the region around where the best constraints lie today. The F4 model corresponds to $|f_{R0}| = 10^{-4}$ (is in tension with local experiments), the F5 model to $|f_{R0}| = 10^{-5}$ (agrees with most experiments and observations, but are in tension with others) and the F6 model to $|f_{R0}| = 10^{-6}$ (which is still allowed). All these models are completely compatible with observations of the expansion history of the Universe, but present large deviations from the $\Lambda$CDM in the clustering and formation of structures.

\bibliographystyle{ws-procs961x669}
\bibliography{refs}

\begin{thebibliography}{10}

\bibitem{Aghanim:2018eyx}
N.~Aghanim {\em et~al.}, {Planck 2018 results. VI. Cosmological parameters}
  (2018).

\bibitem{Khoury:2003aq}
J.~Khoury and A.~Weltman, {Chameleon fields: Awaiting surprises for tests of
  gravity in space}, {\em Phys. Rev. Lett.} {\bf 93}, p. 171104  (2004).

\bibitem{Khoury:2003rn}
J.~Khoury and A.~Weltman, {Chameleon cosmology}, {\em Phys. Rev.} {\bf D69}, p.
  044026  (2004).

\bibitem{Verde:2019ivm}
L.~Verde, T.~Treu and A.~G. Riess, {Tensions between the Early and the Late
  Universe}, {\em Nature Astron.} {\bf 3}, p. 891 (7 2019).

\bibitem{Bertschinger:2006aw}
E.~Bertschinger, {On the Growth of Perturbations as a Test of Dark Energy},
  {\em Astrophys. J.} {\bf 648}, 797  (2006).

\bibitem{Bertschinger:2008zb}
E.~Bertschinger and P.~Zukin, {Distinguishing Modified Gravity from Dark
  Energy}, {\em Phys. Rev.} {\bf D78}, p. 024015  (2008).

\bibitem{Linder:2018jil}
E.~V. Linder, {No Slip Gravity}, {\em JCAP} {\bf 03}, p. 005  (2018).

\bibitem{Starobinsky:1980te}
A.~A. Starobinsky, {A New Type of Isotropic Cosmological Models Without
  Singularity}, {\em Phys. Lett. B} {\bf 91}, 99  (1980).

\bibitem{Carroll:2003wy}
S.~M. Carroll, V.~Duvvuri, M.~Trodden and M.~S. Turner, {Is cosmic speed - up
  due to new gravitational physics?}, {\em Phys. Rev.} {\bf D70}, p. 043528
  (2004).

\bibitem{Capozziello:2002rd}
S.~Capozziello, {Curvature quintessence}, {\em Int. J. Mod. Phys.} {\bf D11},
  483  (2002).

\bibitem{Hu:2007nk}
W.~Hu and I.~Sawicki, {Models of f(R) Cosmic Acceleration that Evade
  Solar-System Tests}, {\em Phys. Rev.} {\bf D76}, p. 064004  (2007).

\bibitem{Magnano:1993bd}
G.~Magnano and L.~M. Sokolowski, {On physical equivalence between nonlinear
  gravity theories and a general relativistic selfgravitating scalar field},
  {\em Phys. Rev. D} {\bf 50}, 5039  (1994).

\bibitem{Jaime:2010kn}
L.~G. Jaime, L.~Patino and M.~Salgado, {Robust approach to f(R) gravity}, {\em
  Phys. Rev. D} {\bf 83}, p. 024039  (2011).

\bibitem{2012JCAP...01..051L}
B.~{Li}, G.-B. {Zhao}, R.~{Teyssier} and K.~{Koyama}, {ECOSMOG: an Efficient
  COde for Simulating MOdified Gravity}, {\em JCAP} {\bf 2012}, p. 051 (Jan
  2012).

\bibitem{Cautun:2017tkc}
M.~Cautun, E.~Paillas, Y.-C. Cai, S.~Bose, J.~Armijo, B.~Li and N.~Padilla,
  {The Santiago–Harvard–Edinburgh–Durham void comparison – I. SHEDding
  light on chameleon gravity tests}, {\em Mon. Not. Roy. Astron. Soc.} {\bf
  476}, 3195  (2018).

\bibitem{White:2016yhs}
M.~White, {A marked correlation function for constraining modified gravity
  models}, {\em JCAP} {\bf 1611}, p. 057  (2016).

\bibitem{Aviles:2019fli}
A.~Aviles, K.~Koyama, J.~L. Cervantes-Cota, H.~A. Winther and B.~Li, {Marked
  correlation functions in perturbation theory}, {\em JCAP} {\bf 01}, p. 006
  (2020).

\bibitem{Alam:2020jdv}
S.~Alam, C.~Arnold, A.~Aviles {\em et~al.}, {Testing the theory of gravity with
  DESI: estimators, predictions and simulation requirements} (11 2020).

\bibitem{Philcox:2020srd}
O.~H.~E. Philcox, A.~Aviles and E.~Massara, {Modeling the Marked Spectrum of
  Matter and Biased Tracers in Real- and Redshift-Space}, {\em JCAP} {\bf 03},
  p. 038  (2021).

\bibitem{Aviles:2017aor}
A.~Aviles and J.~L. Cervantes-Cota, {Lagrangian perturbation theory for
  modified gravity}, {\em Phys. Rev.} {\bf D96}, p. 123526  (2017).

\bibitem{Aviles:2020wme}
A.~Aviles, G.~Valogiannis, M.~A. Rodriguez-Meza, J.~L. Cervantes-Cota, B.~Li
  and R.~Bean, {Redshift space power spectrum beyond Einstein-de Sitter
  kernels}, {\em JCAP} {\bf 04}, p. 039  (2021).

\bibitem{Beisbart:2000ja}
C.~Beisbart and M.~Kerscher, {Luminosity- and morphology-dependent clustering
  of galaxies}, {\em Astrophys. J.} {\bf 545}, p.~6  (2000).

\bibitem{Sheth:2005ai}
R.~K. Sheth, {The halo-model description of marked statistics}, {\em Mon. Not.
  Roy. Astron. Soc.} {\bf 364}, p. 796  (2005).

\bibitem{Sheth:2005aj}
R.~K. Sheth, A.~J. Connolly and R.~Skibba, {Marked correlations in galaxy
  formation models}, {\em Submitted to: Mon. Not. Roy. Astron. Soc.}   (2005).

\bibitem{White:2008ii}
M.~White and N.~Padmanabhan, {Breaking Halo Occupation Degeneracies with Marked
  Statistics}, {\em Mon. Not. Roy. Astron. Soc.} {\bf 395}, p. 2381  (2009).

\bibitem{Valogiannis:2017yxm}
G.~Valogiannis and R.~Bean, {Beyond $\delta$: Tailoring marked statistics to
  reveal modified gravity}, {\em Phys. Rev.} {\bf D97}, p. 023535  (2018).

\bibitem{Hernandez-Aguayo:2018yrp}
C.~Hern\'andez-Aguayo, C.~M. Baugh and B.~Li, {Marked clustering statistics in
  $f(R)$ gravity cosmologies}, {\em Mon. Not. Roy. Astron. Soc.} {\bf 479},
  4824  (2018).

\bibitem{Armijo:2018urs}
J.~Armijo, Y.-C. Cai, N.~Padilla, B.~Li and J.~A. Peacock, {Testing modified
  gravity using a marked correlation function}, {\em Mon. Not. Roy. Astron.
  Soc.} {\bf 478}, 3627  (2018).

\bibitem{Desjacques:2016bnm}
V.~Desjacques, D.~Jeong and F.~Schmidt, {Large-Scale Galaxy Bias}, {\em Phys.
  Rept.} {\bf 733}, 1  (2018).

\bibitem{Aviles:2018saf}
A.~Aviles, M.~A. Rodriguez-Meza, J.~De-Santiago and J.~L. Cervantes-Cota,
  {Nonlinear evolution of initially biased tracers in modified gravity}, {\em
  JCAP} {\bf 1811}, p. 013  (2018).

\bibitem{Gottloeber:2002vm}
S.~Gottloeber, M.~Kerscher, A.~V. Kravtsov, A.~Faltenbacher, A.~Klypin and
  V.~Mueller, {Spatial distribution of galactic halos and their merger
  histories}, {\em Astron. Astrophys.} {\bf 387}, p. 778  (2002).

\bibitem{Skibba:2005kb}
R.~Skibba, R.~K. Sheth, A.~J. Connolly and R.~Scranton, {The
  luminosity-weighted or `marked' correlation function}, {\em Mon. Not. Roy.
  Astron. Soc.} {\bf 369}, 68  (2006).

\bibitem{White:2014gfa}
M.~White, {The Zel'dovich approximation}, {\em Mon. Not. Roy. Astron. Soc.}
  {\bf 439}, 3630  (2014).

\bibitem{Baldauf:2015xfa}
T.~Baldauf, M.~Mirbabayi, M.~Simonović and M.~Zaldarriaga, {Equivalence
  Principle and the Baryon Acoustic Peak}, {\em Phys. Rev.} {\bf D92}, p.
  043514  (2015).

\bibitem{Matsubara:2008wx}
T.~Matsubara, {Nonlinear perturbation theory with halo bias and redshift-space
  distortions via the Lagrangian picture}, {\em Phys. Rev.} {\bf D78}, p.
  083519  (2008), [Erratum: Phys. Rev.D78,109901(2008)].

\bibitem{Aviles:2018thp}
A.~Aviles, {Renormalization of Lagrangian bias via spectral parameters}, {\em
  Phys. Rev.} {\bf D98}, p. 083541  (2018).

\bibitem{Eggemeier:2018qae}
A.~Eggemeier, R.~Scoccimarro and R.~E. Smith, {Bias Loop Corrections to the
  Galaxy Bispectrum}, {\em Phys. Rev.} {\bf D99}, p. 123514  (2019).

\bibitem{Carlson:2012bu}
J.~Carlson, B.~Reid and M.~White, {Convolution Lagrangian perturbation theory
  for biased tracers}, {\em Mon. Not. Roy. Astron. Soc.} {\bf 429}, p. 1674
  (2013).

\bibitem{Vlah:2015sea}
Z.~Vlah, M.~White and A.~Aviles, {A Lagrangian effective field theory}, {\em
  JCAP} {\bf 1509}, p. 014  (2015).

\bibitem{Kaiser:1984sw}
N.~Kaiser, {On the Spatial correlations of Abell clusters}, {\em Astrophys. J.}
  {\bf 284}, L9  (1984).

\bibitem{Mo:1996cn}
H.~J. Mo, Y.~P. Jing and S.~D.~M. White, {High-order correlations of peaks and
  halos: A Step toward understanding galaxy biasing}, {\em Mon. Not. Roy.
  Astron. Soc.} {\bf 284}, p. 189  (1997).

\bibitem{Sheth:1999mn}
R.~K. Sheth and G.~Tormen, {Large scale bias and the peak background split},
  {\em Mon. Not. Roy. Astron. Soc.} {\bf 308}, p. 119  (1999).

\bibitem{Schmidt:2012ys}
F.~Schmidt, D.~Jeong and V.~Desjacques, {Peak-Background Split,
  Renormalization, and Galaxy Clustering}, {\em Phys. Rev.} {\bf D88}, p.
  023515  (2013).

\bibitem{Tassev:2013rta}
S.~Tassev, {Lagrangian or Eulerian; Real or Fourier? Not All Approaches to
  Large-Scale Structure Are Created Equal}, {\em JCAP} {\bf 1406}, p. 008
  (2014).

\bibitem{2013MNRAS.428.1036M}
M.~{Manera}, R.~{Scoccimarro}, W.~J. {Percival}, L.~{Samushia}, C.~K.
  {McBride}, A.~J. {Ross}, R.~K. {Sheth}, M.~{White}, B.~A. {Reid}, A.~G.
  {S{\'a}nchez}, R.~{de Putter}, X.~{Xu}, A.~A. {Berlind}, J.~{Brinkmann},
  C.~{Maraston}, B.~{Nichol}, F.~{Montesano}, N.~{Padmanabhan}, R.~A. {Skibba},
  R.~{Tojeiro} and B.~A. {Weaver}, {The clustering of galaxies in the SDSS-III
  Baryon Oscillation Spectroscopic Survey: a large sample of mock galaxy
  catalogues}, {\em Mon. Not. Roy. Astron. Soc.} {\bf 428}, 1036 (Jan 2013).

\bibitem{2012arXiv1210.1833A}
D.~{Alonso}, {CUTE solutions for two-point correlation functions from large
  cosmological datasets}, {\em arXiv e-prints} , p. arXiv:1210.1833 (Oct 2012).

\bibitem{Massara:2020pli}
E.~Massara, F.~Villaescusa-Navarro, S.~Ho, N.~Dalal and D.~N. Spergel, {Using
  the Marked Power Spectrum to Detect the Signature of Neutrinos in Large-Scale
  Structure}, {\em Phys. Rev. Lett.} {\bf 126}, p. 011301  (2021).

\bibitem{Lesgourgues:2006nd}
J.~Lesgourgues and S.~Pastor, {Massive neutrinos and cosmology}, {\em Phys.
  Rept.} {\bf 429}, 307  (2006).

\bibitem{Philcox:2020fqx}
O.~H.~E. Philcox, E.~Massara and D.~N. Spergel, {What does the marked power
  spectrum measure? Insights from perturbation theory}, {\em Phys. Rev. D} {\bf
  102}, p. 043516  (2020).

\bibitem{Koyama:2015vza}
K.~Koyama, {Cosmological Tests of Modified Gravity}, {\em Rept. Prog. Phys.}
  {\bf 79}, p. 046902  (2016).

\end{thebibliography}

\end{document}